\documentclass{mem}
\usepackage{natbib}
\usepackage{txfonts}
\usepackage{balance}
\usepackage{graphicx}
\usepackage[a4paper,breaklinks,dvipdfm]{hyperref}
\idline{75}{282}
\begin{document}
\def\teff{$T\rm_{eff }$}
\def\kms{$\mathrm {km s}^{-1}$}
\title{Observed properties of red supergiant and massive AGB star populations}
%\subtitle{}
\author{Jacco Th. \,van Loon}
\institute{Lennard-Jones Laboratories, Keele University, ST5 5BG, UK\\
\email{j.t.van.loon@keele.ac.uk}}
\authorrunning{van Loon}
\titlerunning{Red supergiants and massive AGB stars}
\abstract{This brief review describes some of the observed properties of the
populations of massive asymptotic giant branch (AGB) stars and red supergiants
(RSGs) found in nearby galaxies, with a focus on their luminosity functions,
mass-loss rates and dust production. I do this within the context of their
r\^ole as potential supernova (SN) progenitors, and the evolution of SNe and
their remnants. The paper ends with an outlook to the near future, in which
new facilities such as the James Webb Space Telescope offer a step change in
our understanding of the evolution and fate of the coolest massive stars in
the Universe.
}
\maketitle{}
\section{Introduction}

In the final stages of their evolution, stars in the birth mass range
$\sim1$--7 M$_\odot$ ascend the AGB, reaching luminosities of order $10^4$
L$_\odot$ due to hydrogen and helium shell burning; Hot Bottom Burning (HBB)
at the bottom of the convection zone in the most massive AGB stars can further
raise the luminosity, approaching $10^5$ L$_\odot$ \citep{Iben83}. The nuclear
evolution of AGB stars is truncated by mass loss \citep{vanLoon99}, leaving
behind a carbon--oxygen white dwarf. Stars in the birth mass range
$\sim11$--30 M$_\odot$ do not develop core degeneracy and instead become core
helium burning RSGs \citep{Maeder12}. RSGs also evolve along a (short) branch,
and their mass loss may or may not change their appearance \citep{vanLoon99}
before they explode as a core-collapse supernova \citep{Smartt15}. This leaves
a range $\sim7$--11 M$_\odot$ unaccounted for; these are the super-AGB stars,
which behave very much like massive AGB stars but ignite carbon burning in the
core before leaving an oxygen--neon--magnesium white dwarf or undergoing
electron-capture core-collapse \citep{Doherty17}. The actual mass range is
much narrower, but the boundaries are uncertain. Understanding the fate of
super-AGB stars, and recognising them in nature, presents one of the greatest
challenges in contemporary astrophysics.

\section{Do we know any super-AGB star?}

No. Claims are sometimes made, on the basis of surface abundances (rubidium,
lithium) or luminosities (above the classical AGB limit), but we lack a way to
unequivocally distinguish super-AGB stars from massive AGB stars experiencing
HBB. Nonetheless, we should keep a record of candidates for when we have a way
to confirm, or refute, their super-AGB nature. Until then, ``discoveries'' of
super-AGB stars are steeped in controversy.

A fascinating example is the bright Harvard Variable, HV\,2112 in the Small
Magellanic Cloud (SMC). It is a little more luminous than the classical AGB
limit, sometimes as cool as spectral type M7.5, and has a relatively long
pulsation period \citep{Wood83} -- commensurate with being such a large star.
Things got more exciting when \citet{Smith90} detected lithium, which
indicated some level of HBB even if not clinching its status as a super-AGB
star. In that same year, however, \citet{Reid90} showed lithium to be absent.

Nothing much happened, until \citet{Levesque14} proposed that HV\,2112 could
also be a Thorne--$\dot{\rm Z}$ytkow object -- a neutron star which ended up
inside a stellar mantle. This conjecture was immediately put to the test by
\citet{Tout14}, who note the difficulty in distinguishing between the
different origins of similar peculiarities -- the calcium abundance might be
the discriminating feature, in favour of a Thorne--$\dot{\rm Z}$ytkow object.

But does HV\,2122 in fact reside in the SMC, and is a relatively massive star?
\citet{Maccarone16} concluded on the basis of proper motion measurements that
HV\,2112 must instead belong to the Galactic Halo, and be an extrinsic S-type
star upon which the anomalous abundances had been imprinted by a companion AGB
star. This would explain the calcium abundance, which is generally high in the
Halo. The radial velocity, whilst not excluding Halo membership, is however
consistent with that of the SMC \citep{Gonzalez15}. Revisiting the proper
motion, \citet{Worley16} reinstate its SMC membership. The saga continues.

\section{Converting luminosity functions into star formation histories}

As stars of different birth masses attain different luminosities during their
lives as AGB stars, super-AGB stars or RSGs, their luminosity distribution
reflects the star formation history (SFH). However, the evolutionary tracks
are difficult to separate especially if there is a spread in metallicity, and
so the luminosity distribution will be a blend of the luminosity evolution of
stars of a range of birth masses, especially on the AGB. Evolution in
luminosity is curtailed, though, once stars lose mass at very high rates, so
if one could identify stars in that extreme final phase the luminosity
distributions would map much more cleanly onto the SFH. Luckily, these stars
develop strong, long-period variability (LPV) as their atmospheres pulsate and
(help) drive the mass loss \citep{Wood83}.

A method was devised to determine the SFH from the luminosity distribution of
LPVs by \cite{Javadi11b}, and applied to a near-infrared monitoring survey of
LPVs in the Local Group spiral galaxy M\,33 \citep{Javadi11a}. If
$dn^\prime(t)$ is the number of LPVs associated with all stars that formed
within a timespan $dt$ around lookback time $t$, then the star formation rate
around that time is
\begin{equation}
\xi(t)\ =\ \frac{dn^\prime(t)}{\delta t}\,
\frac{\int_{\rm min}^{\rm max}f_{\rm IMF}(m)m\,dm}
{\int_{m(t)}^{m(t+dt)}f_{\rm IMF}(m)\,dm},
\end{equation}
where $f_{\rm IMF}$ is the initial mass ($m$) function and $\delta t$ the LPV
lifetime.

The method relies on stellar evolution models; only the Padova group predicts
all of the essential information \citep{Marigo17}. This is a problem, as
different models predict a different final luminosity for a given birth mass,
and different lifetimes before and during the LPV phase \citep{Marigo17}. It
is also a blessing {\it because} it is model dependent, as it opens an avenue
into calibrating the models and thus offers insight into the physics. This
became apparent very soon, when \citet{Javadi13} found that the inferred
integrated mass loss exceeded the birth mass. If we define
\begin{equation}
\eta\ =\ \frac{\Sigma_{i=1}^N\left(\dot{M}_i\times(\delta t)_i\right)}
{\Sigma_{i=1}^NM_i},
\end{equation}
then $\eta(m)=1-m_{\rm final}/m_{\rm birth}<1$ always, but the measurements
yielded $\eta>1$. At least part of the solution had to be a downward
revision of the LPV lifetimes, which was later confirmed independently by the
Padova group themselves \citep{Rosenfield16}. The revised lifetimes were used
to derive a more reliable SFH for M\,33 \citep{Javadi17}.

\section{Mass loss from dusty AGB stars and red supergiants}

Mass-loss rates can be determined from dusty AGB stars and RSGs by modelling
the spectral energy distribution (SED) using a radiative transfer model. The
first such systematic analysis for stars with known distances was performed
for populations within the Large Magellanic Cloud (LMC) by \citet{vanLoon99},
who found that mass-loss rates exceeded nuclear consumption rates for most of
the AGB evolution but not for much of the RSG evolution. This means that AGB
stars avoid SN but RSGs do not.

They also found evidence that the mass-loss rate increases during the
evolution along the AGB as the luminosity increases and the photosphere cools,
though RSGs in particular seem to exhibit more abrupt variations between mild
and possibly short episodes of much enhanced mass loss. This was put on a more
quantitative footing by \citet{vanLoon05} who parameterised the mass-loss rate
during the dusty phase of (massive) AGB and RSG evolution as
\begin{eqnarray}
\log{\dot{M}}\ =\ & -5.65+1.05\log(L/10\,000\,{\rm L}_\odot) \nonumber \\
                  & -6.3\log(T_{\rm eff}/3500\,{\rm K}),
\end{eqnarray}
with no evidence for a metallicity dependence. The dependence on temperature
seems very strong but it must be realised that the temperatures among these
kinds of stars fall within a limited range ($\sim2500$--4000 K).

Because the mass-loss rate that is derived from the SED depends on how much
the dust is diluted it has a dependence on the wind speed, $v_{\rm exp}$. What
is really seen directly is the optical depth of the dusty envelope:
\begin{equation}
\tau\ \propto\ \frac{\dot{M}}{r_{\rm gd}v_{\rm exp}\sqrt{L}},
\end{equation}
where $r_{\rm gd}$ is the gas:dust mass ratio. However, if the wind is driven
by radiation pressure upon the dust, then
\begin{equation}
v_{\rm exp}\ \propto\ \frac{L^{1/4}}{r_{\rm gd}^{1/2}}.
\end{equation}
By measuring $v_{\rm exp}$ directly, for instance from the double-horned
hydroxyl maser line profile, its value can be reconciled with the SED
modelling by tuning the value for $r_{\rm gd}$. This led \citet{Goldman17} to
determine a more sophisticated formula for the mass-loss rates from AGB stars
and RSGs in the LMC, Galactic Centre and Galactic Bulge with known pulsation
periods $P$:
\begin{eqnarray}
\log{\dot{M}}\ =\ -4.97+0.90\log(L/10\,000\,{\rm L}_\odot) \nonumber \\
                  +0.75\log(P/500\,{\rm d})-0.03\log(r_{\rm gd}/200).
\end{eqnarray}
This time $P$ instead of $T_{\rm eff}$ measures the size of the star (in
combination with $L$), but the resemblance to the formula found earlier is
striking. There is no dependence on the gas:dust ratio and, by inference, on
the metallicity.

The mass loss probed in the dusty phases may only be part of the story. If
sustained over an extended period of time, moderate mass loss may matter too.
Likewise, the metallicity of massive AGB stars and RSGs in the SMC is only
$\sim0.2$ Z$_\odot$ (as opposed to $\sim0.5$ Z$_\odot$ in the LMC), and stars
become dusty later in their evolution, possibly having lost more mass in other
ways before. \citet{Bonanos10} indeed found that most RSGs exhibit moderate
mass-loss rates, $\sim10^{-6}$ M$_\odot$ yr$^{-1}$, which agrees with the study
by \citet{Mauron11} of Galactic RSGs among which only few exhibit mass-loss
rates $>10^{-5}$ M$_\odot$ yr$^{-1}$. While this confirms the earlier findings
by \citet{vanLoon99}, no clear bimodality was seen.

Larger populations of massive AGB stars and RSGs are needed to be more
conclusive about the rarest, but most intense phases of mass loss in
comparison to the more common, gentler phases. To that aim, following from
\citet{Javadi13} and the extended survey by \citet{Javadi15}, Javadi et al.\
(in prep.) have measured mass-loss rates for thousands such stars in M\,33.
They confirm the gradual evolution in mass loss along the AGB and the bimodal
mass loss on the RSG, with the rates increasing in proportion to luminosity.

These studies show no evidence for anything peculiar to be happening to
super-AGB stars. If anything, they are most likely to follow the extension of
the AGB sequence and attain exuberant mass-loss rates of $\sim10^{-4}$
M$_\odot$ yr$^{-1}$. An accurate assessment of the integrated mass loss might
lead us to exclude -- or allow -- their possible fate as electron-capture SN.

An alternative route to mapping the evolution of mass loss along the AGB or
RSG branch is based on the fact that populous clusters may show more than one
such example. Given the fast evolution in those advanced stages, we are
essentially watching a star of the same mass at different moments in its
evolution, as snaphots in a movie. \citet{Davies08} pioneered this approach in
the Galactic cluster RSG\,C1, and more recently \citet{Beasor16} applied the
same principle to the LMC cluster NGC\,2100. They both confirm the increase in
mass loss along the RSG branch. Such studies in clusters in which we {\it
know} super-AGB stars should form may elucidate the properties, evolution and
fate of super-AGB stars where field studies struggle to recognise them. Apart
from the difficulty in finding such clusters, it may be difficult to catch
them at their most extreme.

\section{Beyond the red supergiant and asymptotic giant branches}

Massive AGB stars and RSGs can undergo a blueward excursion as a result of
core expansion before returning to the cool giant branch -- a ``blue loop''.
This is a much slower transformation than the ``jump'' from the main sequence
to the giant branch, and is responsible for populating the Cepheid instability
strip \citep{Valle09}. Cepheid variables therefore could be extremely valuable
probes of what may already have happened to these stars on the cool giant
branch, such as any mass loss, especially as their pulsational properties
depend on their current mass. It also means that the cool giant branches are
populated by stars that have, and those that have not undergone a blue loop --
again, these may differ as a result of the mass loss during their lives as
blue (or yellow) supergiants. Typically, much fewer warm supergiants are seen
than are predicted by the models \citep{Neugent10}, which suggests that the
models do not yet adequately account for certain processes that happen inside
RSGs.

Pulsation periods lengthen as a star expands when it is reduced in mass, so
period--luminosity diagrams of AGB and RSG LPVs have the potential to trace
mass loss and possibly the effects of blue loops. Also, the mode in which the
pulsation is excited depends on stellar structure. \citet{Yang12} charted this
parameterspace for RSGs in the Magellanic Clouds, M\,33 and the Milky Way but
their combination with similar information for AGB stars left an unfortunate
gap right around where super-AGB stars could be found. This must -- and can --
be remedied.

AGB stars and RSGs may also move irreversibly towards hotter photospheres, due
to mass loss (by a wind or stripping in a binary system). The luminosity
distribution over post-AGB stars and post-RSGs must be devoid of the birth
masses associated with the electron-capture SN demise of the more massive
among super-AGB stars, and depleted of those RSGs which encountered their end
in a SN. Again, a concerted analysis is required, where in the past different
communities have often concentrated on just the lower-mass (post-)AGB stars or
on just the higher-mass (post-)RSGs.

\begin{figure*}[t!]
\resizebox{\hsize}{!}{
\begin{tabular}{ll}
\includegraphics[clip=true]{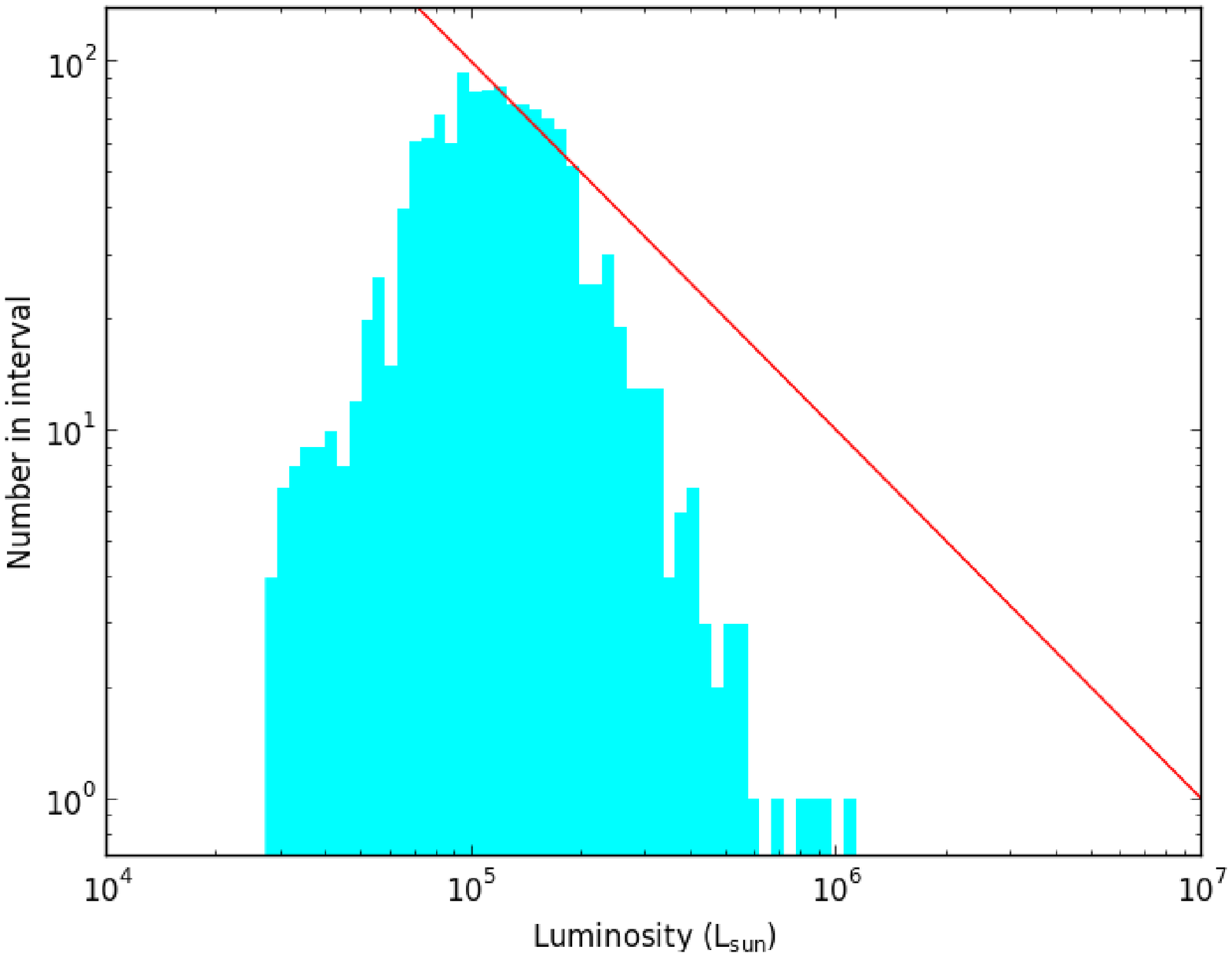} &
\includegraphics[clip=true]{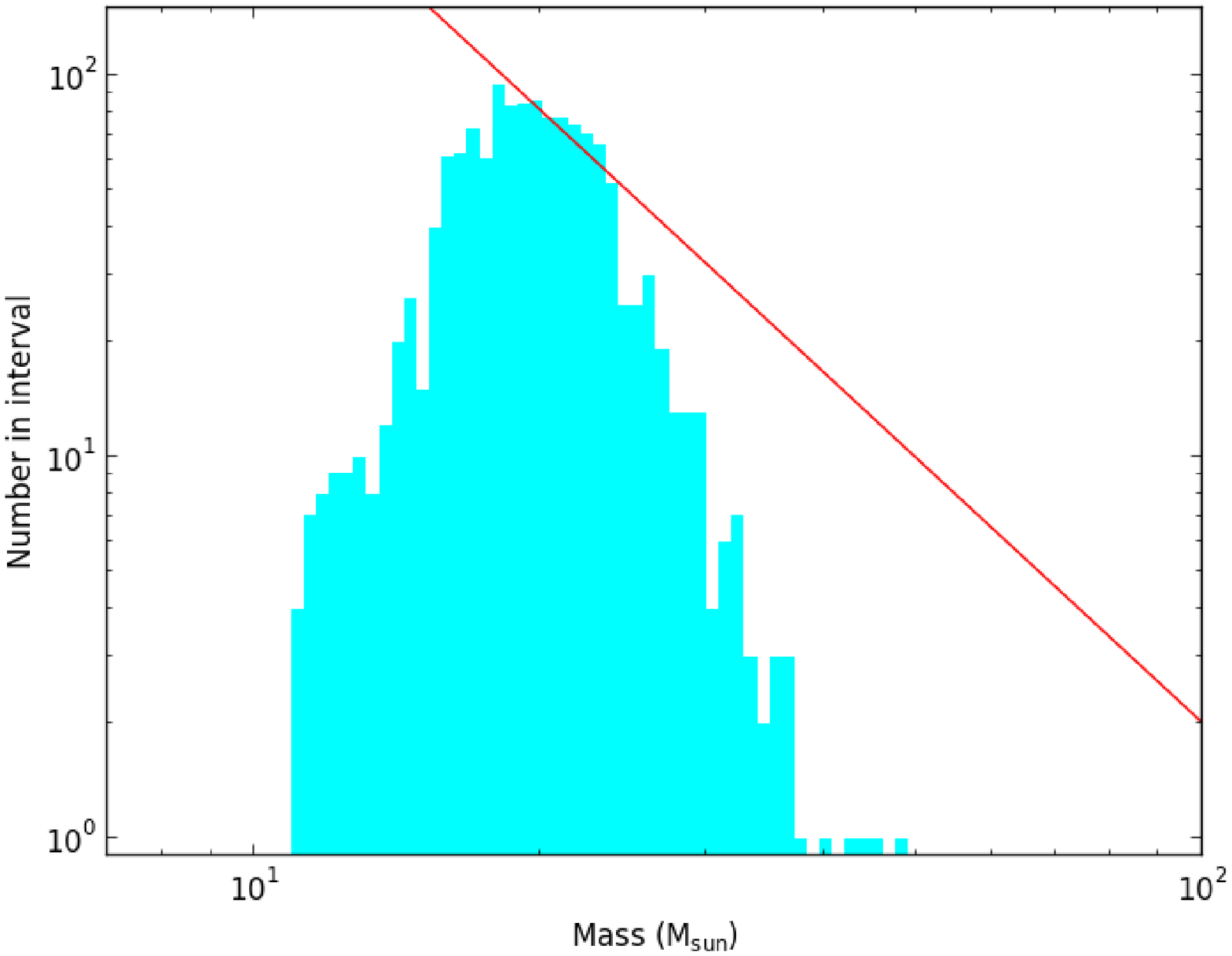}
\end{tabular}
}
\caption{\footnotesize
Luminosity (left) and birth mass (right) distribution of RSGs in the
grand-design spiral galaxy M\,101 at 7 Mpc distance based on groundbased and
{\it Spitzer}-IRAC infrared imaging (courtesy of James Bamber). The
luminosity-to-mass conversion is based on a fit to Padova models
\citep{Marigo17}: $\log L = 2.49\log M + 1.84$. The initial mass function,
with a slope of $-2.3$, is overplotted; the distributions are depleted on the
low and high sides due to incompleteness and diminished RSG lifetimes,
respectively.}
\label{m101}
\end{figure*}

\section{The progenitors and remnants of core-collapse supernov{\ae}}

Having established that many RSGs most likely do not lose their envelope
before the core collapses, it is comforting that all of the SN type II-P
progenitors that have been discovered so far are RSGs \citep{Smartt15}. Their
birth mass range is estimated to be $\sim9$--17 M$_\odot$ \citep{Smartt15},
though the upper limits could allow dusty RSGs as massive as $\sim21$
M$_\odot$ to have resulted in II-P SNe. This would also be more consistent
with the above findings about the mass-loss rates of RSGs, where a proportion
of core-collapsing RSGs should experience high mass-loss rates. The confirmed
RSG progenitors tend to be of relatively early spectral type and thus
constitute those RSGs that have lost relatively little of their mantle.

What exactly determines the difference in rate of evolution of the core --
setting the timing of core collapse -- and of the mantle -- setting the mass
loss, is unclear, but larger samples of discoveries and limits on progenitors
of II-P SNe should help elucidate this: their lightcurves and spectral
evolution may reveal differences in the mantle mass and circumstellar density,
whilst their galactic environments may reveal a dependence on metallicity.
Likewise, a connection between RSG evolution and mass loss on the one hand,
and SNe of types II-L and Ib on the other, may be made if the more massive
RSGs ($\sim20$--30 M$_\odot$ lose most of their mantle before exploding.

A metallicity effect on the properties of the SN resulting from the explosion
of a RSG (or super-AGB star) is expected even if the mass-loss rate does not
change, because the winds are slower at lower metallicity \citep{Goldman17}.
This would increase the circumburst density, with implications for the SN
lightcurve \citep{Moriya17}. It also means that the reverse shock may be
stronger.

The SN has a devastating effect on the dust produced by the RSG, even if it
did not explode as a RSG as the SN ejecta move at $\sim10^3$ times the speed
of the RSG wind and catch up within a century. The RSG dust, and any ISM dust
in the vicinity, is sputtered in the forward shock, whilst the reverse shock
sputters much of the dust produced in the SN ejecta. The net effect being that
SNe are dust destroyers \citep{Lakicevic15,Temim15}.

Because SN remnants (SNRs) can be seen for $>10^4$ yr they might tell us
something about the SN progenitors in their final $<10^4$ yr that is difficult
to capture while they are still alive. The two most prolific SN factories
known, NGC\,6946 \citep{Sugerman12} and M\,83 \citep{Blair15} exhibit hundreds
of SNRs \citep{Bruursema14,Winkler17}.

\section{The next frontier}

A naive estimate for a massive spiral galaxy suggests one RSG explodes every
century. With a typical RSG lifetime of $\sim10^3$ centuries, we thus expect
to find a population of order $10^3$ RSGs. Thus the statistics look extremely
promising for studies of the luminosity distribution and the evolution of mass
loss of RSGs (and super-AGB stars) in such galaxies.

The next step from the Magellanic Clouds can take us to the nearest spiral
galaxies, M\,33 \citep{Drout12} and M\,31 \citep{Massey16} at $\sim0.9$ Mpc,
or NGC\,300 and the metal-poor dwarf Sextans A, both within 2 Mpc. Indeed, the
Surveying the Agents of Galaxy Evolution (SAGE) team, who have revolutionised
our views of the Magellanic Clouds, are proposing an Early Release Science
programme for the James Webb Space Telescope (JWST) precisely to do that. But
how far could we go?

M\,101 is the most massive spiral galaxy within 7 Mpc, viewed face on. Bamber
et al.\ (in prep.) have used groundbased and {\it Spitzer} infrared images,
and {\it Hubble} optical images, to identify RSG candidates across the entire
disc. The {\it Spitzer} data are heavily compromised by the limits in
resolution and sensitivity, and JWST will be both necessary and sufficient to
quantify the dust production by RSGs in M\,101. Likewise, the luminosity
distribution suffers from incompleteness at the low end. Still, it reveals a
tantalising first glimpse of the evolution of the most luminous RSGs
(Fig.~\ref{m101}): a healthy number of RSGs in the $\sim17$--22 M$_\odot$
range suggests these could well be the progenitors of SNe, while the sharp
drop that sets in at higher masses suggests much diminished RSG lifetimes. The
latter is not unexpected if those are the RSGs that become Wolf--Rayet stars,
as a result of stronger mass loss.

\section{Will we find super-AGB stars?}

Yes. We look towards our colleagues who model the structure and dynamical
behaviour, and nucleosynthesis and surface enrichment of super-AGB stars to
make predictions for the observable signatures that can tell super-AGB stars
apart from other massive AGB stars (and RSGs). We also need to reach an
agreement on the birth-mass range of super-AGB stars. Meanwhile, we look for
peculiarities or more subtle hints of deviations, that indicate we may be
dealing with a star of a different kind.

\begin{acknowledgements}
I would like to warmly thank Paolo Ventura, Flavia Dell'Agli and Marcella Di
Criscienzo and all participants for an interesting and pleasant meeting, my
collaborators and students -- in particular Atefeh Javadi, Steven Goldman and
James Bamber some of whose results I presented here -- and two new feline
friends for their affection on my walk up to the observatory.
\end{acknowledgements}
\bibliographystyle{aa}

\end{document}